\documentclass[11pt,a4paper]{article}
\pdfoutput=1
\usepackage{jcappub}
\usepackage{xcolor}
\usepackage{graphicx}
\usepackage{float}
\usepackage{wrapfig}
\usepackage{bm}
\usepackage{ulem}

\usepackage{amsmath,amssymb}
\usepackage{subcaption}
\usepackage{verbatim}
\usepackage{makeidx}
\usepackage{braket}
\usepackage{multirow}
\usepackage[utf8]{inputenc}

\definecolor{red}{rgb}{1,0,0}
\definecolor{darkred}{rgb}{0.6,0,0}
\definecolor{darkgreen}{rgb}{0.992447,0.623778,0.034597}
\definecolor{ppink}{rgb}{1,0.4,0.4}
\definecolor{bblue}{rgb}{0.284602,0.317763,0.963947}

\newcommand{\vev}[1]{ \left< {#1} \right> }

\newcommand{\dd}{\mathrm{d}}

\newcommand{\R}{\text{R} }
\newcommand{\GW}{\text{GW} }

\makeatletter
\newcommand\footnoteref[1]{\protected@xdef\@thefnmark{\ref{#1}}\@footnotemark}
\makeatother

\allowdisplaybreaks[1]

\title{Gravitational Waves Induced by Scalar Perturbations during a Gradual Transition from an Early Matter Era to the Radiation Era}

\author[a,b]{Keisuke Inomata}
\affiliation[a]{ICRR, The University of Tokyo, Kashiwa, Chiba 277-8582, Japan}
\affiliation[b]{Kavli IPMU (WPI), UTIAS, The University of Tokyo, Kashiwa, Chiba 277-8583, Japan}

\author[c,d]{Kazunori Kohri}
\affiliation[c]{Theory Center, IPNS, KEK, 1-1 Oho, Tsukuba, Ibaraki 305-0801, Japan}
\affiliation[d]{The Graduate University for Advanced Studies (SOKENDAI), 1-1 Oho, Tsukuba, Ibaraki 305-0801, Japan}

\author[e]{Tomohiro Nakama}
\affiliation[e]{Institute for Advanced Study, The Hong Kong University of Science and Technology, Clear Water Bay, Kowloon, Hong Kong}

\author[c]{Takahiro Terada}

\abstract{
We revisit the effects of an early matter-dominated era on gravitational waves induced by scalar perturbations. We carefully take into account the evolution of the gravitational potential, the source of these induced gravitational waves, during a gradual transition from an early matter-dominated era to the radiation-dominated era, where the transition timescale is comparable to the Hubble time at that time. 
Realizations of such a gradual transition include the standard perturbative reheating with a constant decay rate.
Contrary to previous works, we find that the presence of an early matter-dominated era does not necessarily enhance the induced gravitational waves due to the decay of the gravitational potential around the transition from an early matter-dominated era to the radiation-dominated era.
}

\begin{document}
\begin{flushright}
IPMU 19-0066,
KEK-TH-2121,
KEK-Cosmo-236
\end{flushright}
\maketitle

\section{Introduction}
\label{sec:intro}

Stochastic gravitational-wave (GW) background is one of the hottest topics in
astrophysics and cosmology.  It is generated from numerous astrophysical processes, such as
mergers of black holes, neutron stars and white dwarfs (see
Ref.~\cite{Christensen:2018iqi} and references therein).  In addition, there can also be cosmological origins such as vacuum
quantum fluctuations during inflation, cosmic strings and scalar
perturbations at the second order~\cite{Ananda:2006af,Baumann:2007zm,Saito:2008jc,Saito:2009jt,Alabidi:2012ex}
(see also Ref.~\cite{Caprini:2018mtu} for a recent review and Ref.~\cite{Kuroyanagi:2018csn} for detection prospects).  In
particular, GWs induced by scalar perturbations at the second order have recently attracted a lot of
attention~\cite{Espinosa:2018eve,Kohri:2018awv,Cai:2018dig,Bartolo:2018evs,Bartolo:2018rku,Unal:2018yaa,Byrnes:2018txb,Inomata:2018epa,Clesse:2018ogk,Cai:2019amo,Cai:2019jah,Wang:2019kaf,Ben-Dayan:2019gll,Tada:2019amh}. 
Such induced GWs can be especially important in inflationary scenarios where small-scale primordial fluctuations are enhanced, to the extent that a cosmologically relevant amount of primordial black holes are created by the gravitational collapse of extremely rare peaks during the radiation-dominated (RD) era. 
In such scenarios, the scalar perturbations may
simultaneously induce stochastic GWs that can be detected by current
and future GW
observations~\cite{Saito:2008jc,Saito:2009jt,Inomata:2016rbd,Garcia-Bellido:2017aan}.
Even if no sizable amount of primordial black holes was produced, the
induced GWs can be used to constrain the amplitude of primordial
perturbations on small
scales~\cite{Inomata:2018epa,Ben-Dayan:2019gll}.  In this and an
accompanying paper~\cite{Inomata:2019ivs}, we argue that the induced GWs could also be useful
probes of an early epoch after inflation and before the RD era.

Throughout this work, we discuss the GWs induced by the scalar (namely
curvature/density) perturbations assuming that an early matter-dominated (eMD) era follows inflation, which ends with the Universe
dominated by radiation
(reheating)~\cite{Assadullahi:2009nf,Alabidi:2013wtp,Kohri:2018awv}. 
After the inflation era, a massive field, such as the inflaton or a
curvaton, 
could dominate the energy density and the Universe behaves as a matter-dominated (MD) era until the field decays to radiation, which is the
beginning of the RD era.  Unlike during the RD era, the gravitational
potential, which is the source of the induced GWs, does not decay even
on subhorizon scales during a MD era. Therefore the presence of an eMD
era which precedes the RD era may seem to enhance the induced GWs, relative to the GWs induced from the same primordial spectrum when the transition from inflation to the
RD era is virtually instantaneous.

The effects of an eMD era on induced GWs were also discussed in
Refs.~\cite{Assadullahi:2009nf,Alabidi:2013wtp,Kohri:2018awv}.  
In the previous works, they 
assumed that the gravitational potential is constant on subhorizon
scales during an eMD era up to the moment of reheating. However, in
realistic situations, the transition from an eMD era to the RD era
occurs gradually due to an exponential decay of the massive particle
($\rho_{\rm m} a^3 \propto \exp(- \Gamma t)$ with
$\Gamma$ denoting its decay rate). As a result, the gravitational potential also changes
gradually.  Here, by a gradual-reheating transition
we mean that the transition timescale is comparable to 
the Hubble time at that time. In
addition, the authors in the previous works neglected an important part of contributions to
the induced GWs associated with the RD era. The part of the contributions mainly
comes from the modes that had entered the horizon during an eMD era and
have non-negligible impacts on the GW spectrum as we show later.  In
short, we take into account the gradual evolution of the gravitational
potential and calculate GWs induced throughout the transition for the
first time.\footnote{The effects of the decrease in the gravitational potential
  during a gradual transition and the resultant incomplete enhancement of induced
  GWs were also discussed in a talk by
  S.~Kuroyanagi~\cite{Kuroyanagi2017talk}, though they neglected the difference between the evolutions of the perturbations of matter and radiation. Some qualitative features of our results including the shape of the spectrum $\Omega_{\text{GW}}$ are somewhat
  different from theirs.}
Note that, throughout this paper, we focus on the GWs induced by the perturbations entering the horizon during an eMD era because the effects of an eMD era on induced GWs are mainly due to the behaviors of the perturbations on subhorizon scales during an eMD era, which are different from those during the RD era.

This paper is organized as follows.  In Sec.~\ref{sec:formalism},
we review the calculations of the induced GWs with some refinements of
previous works. 
The gradual reheating and its consequences, particularly the
significant suppression of the gravitational potential, are discussed
in Sec.~\ref{sec:phi_evo}.  Using the results obtained in these
sections, we compute the induced GWs in Sec.~\ref{sec:gw_calculation}.
Sec.~\ref{sec:discussion} is dedicated to discussions.

\section{A formalism to calculate induced gravitational waves}
\label{sec:formalism}

In this section, we introduce formulas to calculate GWs induced around a transition from an eMD era to the RD era. 
We assume that the curvature perturbations follow a Gaussian
distribution.\footnote{GWs induced by scalar perturbations following a
  non-Gaussian distribution are studied in
  Refs.~\cite{Nakama:2016gzw,Garcia-Bellido:2017aan,Cai:2018dig,Unal:2018yaa}.} 
We take the conformal Newtonian gauge,\footnote{ The gauge dependence
  of induced GWs is discussed in Refs.~\cite{Hwang:2017oxa,
    Arroja:2009sh}.}  and the metric perturbations are given by
\begin{align}
    \dd s^2 =& a^2 \left( -(1+2\Phi) \dd \eta^2  + \left((1-2\Psi) \delta_{ij} + \frac{1}{2} h_{ij} \right) \dd x^i \dd x^j \right),   \label{eq:metric}
\end{align}
where $\eta = \int dt/a(t)$ is the cosmic conformal time  with $a$ denoting the
scale factor. Since we are interested in the second-order contributions
to the tensor perturbation $h_{ij}$ originating from the first-order
scalar perturbations $\Phi$ and $\Psi$, we neglect the irrelevant
components such as the vector perturbations and first-order tensor
perturbations.  We also neglect anisotropic stress and take
$\Psi = \Phi$ in the following.  
Note that since $\Phi$ can be interpreted as the Newtonian potential, $\Phi$ is referred to as ``gravitational potential''.
Throughout this work, we focus on
small-scale perturbations with
$k \gg k_\text{eq} = 0.0103\, \text{Mpc}^{-1}$~\cite{Aghanim:2018eyx}
and do not take into account the effects associated with the late MD
era ($z\lesssim 3400$)~\cite{Mollerach:2003nq, Baumann:2007zm}. 
Here, we introduce the formulas only in the case where the transition
from an eMD era to the RD era suddenly occurs at a conformal time
$\eta = \eta_\R$.  In Sec.~\ref{sec:phi_evo}, we will explain how
to take into account a gradual transition.  Since the energy density
is continuous at the transition, the scale factor $a=a(\eta)$ and its
derivative with respect to the conformal time are also continuous.
Then, we can express the scale factor as
\begin{align}
\frac{a(\eta)}{a(\eta_\R)} = 
\begin{cases}
\left( \frac{\eta}{\eta_\R} \right)^2 &\ (\eta < \eta_\R), \\
2 \frac{\eta}{\eta_\R}  -1  &\ (\eta \geq \eta_\R).
\end{cases}
\label{eq:a_app}
\end{align}
\\
Tensor perturbations can be expanded with the Fourier components as 
\begin{align}
h_{ij}(\eta, \bm x) = \int \frac{\dd^3 k}{(2\pi)^{3/2}} \left( e_{ij}^{+}(\bm k) h_{\bm k}^+ + e_{ij}^\times(\bm k) h_{\bm k}^\times \right) e^{i \bm k \cdot \bm x},
\label{eq:h_fourier}
\end{align}
where $e^{\lambda}_{ij}$ $(\lambda : +, \times)$ are the polarization tensors.
The power spectrum of tensor perturbations is defined as
\begin{align}
\vev{h^\lambda_{\bm k}(\eta) h^{\lambda'}_{\bm k'}(\eta)} = \delta_{\lambda \lambda'} \delta^3(\bm k + \bm k') \frac{2\pi^2}{k^3} \mathcal P_h(\eta, k),
\end{align}
where the angle brakets denote an ensemble average.
The energy density parameter of GWs per logarithmic interval in $k$ is given by
\begin{align}
\Omega_{\rm{GW}} (\eta, k) &= \frac{\rho_{\rm{GW}} (\eta, k) } { \rho_{\rm{tot}}(\eta)} \nonumber\\ 
&=
\frac{1}{24} \left( \frac{k}{\mathcal H(\eta) } \right)^2 
\overline{\mathcal P_{h} (\eta,k)} , 
\label{eq:gw_formula}
\end{align}
\\
where $\rho$ represents the energy density, $\mathcal H=a'/a$, with the prime representing derivative with respect to the conformal time, and the overline on the power spectrum indicates the time average over oscillations.
The equation of motion of the tensor perturbations is given by~\cite{Baumann:2007zm}
\begin{align}
h^{\lambda''}_{\bm{k}}(\eta) + 2 \mathcal H h^{\lambda'}_{\bm{k}}(\eta) + k^2 h^{\lambda}_{\bm{k}}(\eta) = 4 S^{\lambda}_{\bm{k}}(\eta).
\label{eq:h_eom}
\end{align}
The source term $S^\lambda_{\bm k}$ is expressed in terms of the Fourier components of the gravitational potential as
\begin{align}
S^\lambda_{\bm{k}} = \int \frac{\dd^3 q}{(2\pi)^{3/2}}e^{\lambda}_{ij}(\bm{k})q_i q_j \left( 2 \Phi_{\bm{q}} \Phi_{\bm{k-q}} + \frac{4}{3(1+w)} (\mathcal H^{-1} \Phi'_{\bm{q}} + \Phi_{\bm{q}}) (\mathcal H^{-1} \Phi'_{\bm{k-q}} + \Phi_{\bm{k-q}})  \right),
\end{align}
where $w = P/\rho$ is the equation-of-state parameter with $P$ being the pressure.
Using the Green's function method, we can express the power spectrum of GWs as~\cite{Kohri:2018awv} (see also~\cite{Ananda:2006af,Baumann:2007zm,Inomata:2016rbd} for detail)
\begin{align}
\overline{\mathcal P_{h} (\eta,k)} = 4\int^\infty_0 &\dd  v \int^{1+v}_{|1-v|} \dd u \left( \frac{4v^2 - (1+v^2 - u^2)^2}{4vu}  \right)^2 \overline{I^2(u,v,k,\eta,\eta_\R) }\mathcal P_\zeta(u k) \mathcal P_\zeta(v k).
\label{eq:p_h_formula}
\end{align}
Here, $\mathcal{P}_\zeta (k)$ is the power spectrum of the primordial
curvature perturbations, the integration variables $u$ and $v$
represent wavenumbers in units of $k$, and the function
$I(u,v,k,\eta,\eta_\R)$ contains information on the dynamics of the
scalar and tensor perturbations and is given by
\begin{align}
I(u,v,k,\eta,\eta_\R) =& \int^{x}_0 \dd \bar{x} \frac{a(\bar \eta)}{a(\eta)} k G_k(\eta, \bar \eta) f(u,v,\bar x, x_\R),
\label{eq:i_formula}
\end{align}
where $x$ and $x_\text{R}$ are defined as $x \equiv k\eta$ and
$x_\text{R} \equiv k\eta_\text{R}$.  In the above expression,
$f(u,v,\bar x, x_\R)$ is the source function defined as
\begin{align}
f(u,v,\bar{x}, x_\R)=& \frac{3 \left( 2(5+3w) \Phi(u\bar{x})\Phi(v\bar{x})+4  \mathcal H^{-1} (\Phi'(u\bar{x})\Phi(v\bar{x})  +  \Phi(u \bar{x})\Phi'(v\bar{x})) 
+ 4 \mathcal H^{-2}  \Phi'(u\bar{x})\Phi'(v\bar{x}) \right) }{25(1+w)}.
\label{eq:f_def}
\end{align}
$\Phi(x,x_\R)$ is the transfer function of the
gravitational potential, which satisfies $\Phi(x\rightarrow 0,x_\R) =
1$. 
The second argument of $\Phi$ is abbreviated in Eq.~\eqref{eq:f_def}
for compact notation, that is, $\Phi(u \bar{x})$ actually means
$\Phi (u\bar{x}, u x_\text{R})$ and $\Phi(v\bar{x})$ should be
understood similarly. The prime here denotes a differentiation
with respect to $\bar{\eta}$.
Since $a$ is given in Eq.~(\ref{eq:a_app}), $k \mathcal H^{-1}$ can be
written as
\begin{align}
k \mathcal H^{-1}(\eta) = 
\begin{cases}
\frac{x}{2} &\ (\eta < \eta_\R), \\
x - \frac{1}{2} x_\R  &\ (\eta \geq \eta_\R).
\end{cases}
\label{eq:h_app}
\end{align}
$G_k$ in Eq.~(\ref{eq:i_formula}) is the Green's function being the solution of
\begin{align}
G''_k(\eta, \bar \eta) + \left( k^2 - \frac{a''(\eta)}{a(\eta)} \right) G_k(\eta, \bar \eta) = \delta (\eta - \bar \eta).
\label{eq:g_formula}
\end{align}
Here, the prime denotes a differentiation with respect to only $\eta$,
not $\bar{\eta}$.  $G_k$ represents the solution for GWs (multiplied
by the scale factor) in the presence of a delta-function source.

Ultimately, the time evolutions of all relevant modes for both $G_k$
and $\Phi$ need to be calculated for gradual-transition scenarios to
obtain induced GWs. Unfortunately, however, we have not been able to
find exact analytic solutions of $G_k$ and $\Phi$ for such
scenarios. In addition, obtaining numerical solutions and plugging
them into the formula for the induced GWs to do the relevant
integrations require high computational costs. Hence, in this paper,
we try to calculate the induced GWs approximately as a first step toward a
more rigorous analysis. Let us begin by using Eq.~(\ref{eq:a_app}) to
decompose Eq.~(\ref{eq:i_formula}) as~\cite{Kohri:2018awv},\footnote{
  Eq.~(\ref{eq:i_formula_emd1}) refines the relevant formula in
  Ref.~\cite{Kohri:2018awv}.  In the limit of $x\rightarrow \infty$,
  the contribution of the first term to $\Omega_\GW$ decreases by
  $1/4$ compared to the previous result.
  \label{fn:footnotename}
}
\begin{align}
\label{eq:i_formula_emd1}
I(u,v,x,x_\R) =& \int^{x_\text{R}}_0 \dd \bar{x} \left( \frac{1}{2(x/x_\R) -1} \right) \left( \frac{\bar x}{x_\text{R}} \right)^2  k G_k^{\text{eMD}\rightarrow \text{RD}}(\eta, \bar \eta) f(u,v,\bar x,x_\R) \nonumber\\
&
+ \int^x_{x_\text{R}} \dd \bar x \left( \frac{2(\bar x/x_\R) -1}{2(x/x_\R) -1} \right) 
k G^{\text{RD}}_k (\eta, \bar \eta) f(u,v,\bar x,x_\R) \\
\label{eq:i_formula_emd}
\equiv& I_\text{eMD}(u,v,x,x_\R) + I_\text{RD}(u,v,x,x_\R),
\end{align}
where we have assumed $x > x_\R$ (see Eq.~(\ref{eq:i_formula_emdonly})
for $x < x_\R$).  The first term $I_\text{eMD}$ represents the
contributions from the GWs induced during an eMD era and propagating
freely during the RD era.  On the other hand, the second term
$I_\text{RD}$ describes the GWs induced during the RD era.  The
Green's functions $G_k$ for GWs are given by
\begin{align}
k G_k^{\text{eMD} \rightarrow \text{RD}} (\eta, \bar \eta) &= C(x, x_\R) \bar x j_1 (\bar x) + D(x, x_\R) \bar x y_1(\bar x), 
\end{align}
and
\begin{align}
k G_k^{\text{RD}} (\eta, \bar \eta) &= \sin (x - \bar x), 
\end{align}
where $j_1$ and $y_1$ are the first and second spherical Bessel
functions, and the coefficients are 
\begin{align}
\label{eq:c_def}
&C(x,x_\R) = \frac{\sin x - 2 x_\R (\cos x + x_\R \sin x) + \sin(x-2x_\R)}{2 x_\R^2}, 
\end{align}
and
\begin{align}
\label{eq:d_def}
&D(x, x_\R) = \frac{(2 x_\R^2 -1) \cos x -2 x_\R \sin x + \cos (x-2 x_\R)}{2x_\R^2}.
\end{align}
Note that when we derive $C$ and $D$, we have connected the GW solutions at the transition requiring continuity of themselves and also their first derivatives~\cite{Kohri:2018awv}.

As we can see from Eqs.~(\ref{eq:f_def}) and (\ref{eq:i_formula_emd1}),
the induced GWs sensitively depend on the evolution of the
gravitational potential $\Phi$.  In
Refs.~\cite{Assadullahi:2009nf,Alabidi:2013wtp}, they assume that
$\Phi$ remains unity until $\eta_\R$ and that $I_\text{RD}$,
representing the contributions from the RD era, is subdominant and
hence can be neglected.  However, as we will see in the next section,
$\Phi$ gradually changes around the transition and therefore we need
to take into account the evolution of $\Phi$ more carefully. The
contributions from the RD era also turn out to have non-negligible
impacts.  These are the main issues we address in this work.

From the above splitting of the function $I$, we have
\begin{align}
\overline {I^2(u,v,x,x_\R)} =& \overline{I_\text{eMD}^2(u,v,x,x_\R)} + \overline{I_\text{RD}^2(u,v,x,x_\R)} + 2\overline{I_\text{RD} I_\text{eMD}(u,v,x,x_\R)}.
\label{eq:i_sep}
\end{align}
For later convenience, we also split $\Omega_\text{GW}$ into three parts as $\Omega_\text{GW} = \Omega_\text{GW,RD} + \Omega_\text{GW,eMD} + \Omega_\text{GW,cross}$, 
where $\Omega_\text{GW,RD}$, $ \Omega_\text{GW,eMD}$, and $\Omega_\text{GW,cross}$ are calculated from $\overline{I^2_\text{RD}}$, $\overline{I^2_\text{eMD}}$ and $2\overline{I_\text{RD} I_\text{eMD}}$, respectively.
Note that $\Omega_\text{GW,cross}$ can be negative.

After the transition, the gravitational potential continues to decay on subhorizon scales and $\Omega_{\text{GW}}$ reaches some constant value.
Hence, we define $\eta_c$ as the moment when $\Omega_\text{GW}$ stops growing. 
Since we consider small-scale perturbations, $\eta_c$ is well before the late matter-radiation equality time.

\section{Evolution of gravitational potential}
\label{sec:phi_evo}

In this section, we focus on the evolution of $\Phi$ around the transition, which has large impacts on the resultant GWs.
To be concrete, we assume that the field that dominates the Universe during an eMD era, such as inflaton, decays to radiation with a constant decay rate $\Gamma$.
In this case, we can use formulas for perturbations in decaying dark matter scenarios~\cite{Ichiki:2004vi,Audren:2014bca,Poulin:2016nat}.\footnote{
Strictly speaking, the perturbations of the coherently oscillating scalar field behave differently from the dust-like fluid for $k \gtrsim \sqrt{a m \mathcal H}$ ($m$: mass of the oscillating field) (see Ref.~\cite{Cembranos:2015oya} and references therein).
	However, for the perturbations that enter the horizon during an eMD era, which we focus on in this paper, we can regard the fluid of the oscillating field as a dust-like fluid (even for the perturbations of the oscillating field).
	The reason is as follows. 
	The comoving horizon scale at the beginning of the eMD era is given as $\sim (a m)^{-1}$ at that time because the inflaton or curvaton starts to oscillate and the eMD era begins when $\mathcal H \sim a m$.
	Then, the wavenumber of the perturbation entering the horizon during the eMD era satisfies $k < m a|_{t = t_\text{eMD,start}} < \sqrt{a m \mathcal H}$ ($t_\text{eMD,start}$: the start time of the eMD era), considering the fact that $\sqrt{am\mathcal H}$ is proportional to $a^{1/4}$ during the eMD era.
	Therefore, we regard the fluid of the oscillating field as a dust-like fluid throughout this paper.
}
Then, the evolutions of the energy densities are described by~\cite{Poulin:2016nat}
\begin{align} 
	\rho'_\text{m} &= -(3\mathcal H + a \Gamma) \rho_\text{m}, \\
	\rho'_\text{r} &= -4\mathcal H \rho_\text{r} + a \Gamma \rho_\text{m},
\end{align}
where 
the subscripts ``m'' and ``r'' represent matter and radiation, respectively.
The equations for perturbations in Fourier space are given by~\cite{Poulin:2016nat}
\begin{align}
	\delta'_\text{m} &= -\theta_\text{m} + 3\Phi' - a\Gamma \Phi, \label{eq:delta_m} \\ 
	\theta'_\text{m} &= -\mathcal H \theta_\text{m} + k^2 \Phi, \label{eq:theta_m} \\	
	\delta'_\text{r} &= -\frac{4}{3} (\theta_\text{r} -3 \Phi') + a\Gamma \frac{\rho_\text{m}}{\rho_\text{r}} ( \delta_\text{m} - \delta_\text{r} + \Phi), \\ 
	\theta'_\text{r} &= \frac{k^2}{4} \delta_\text{r} + k^2 \Phi - a\Gamma \frac{3\rho_\text{m}}{4\rho_\text{r}} \left( \frac{4}{3} \theta_\text{r} - \theta_\text{m} \right) \label{eq:theta_r},
\end{align}
where $\delta$ and $\theta$ denote the energy density perturbation and the velocity divergence~\cite{Poulin:2016nat,Ma:1995ey}, respectively, and we have neglected the anisotropic stress of radiation.
In addition, the derivative of $\Phi$ is given by~\cite{Mukhanov:991646}, 
\begin{align}
	\Phi' = - \frac{k^2 \Phi + 3 \mathcal{H}^2 \Phi + \frac{3}{2} \mathcal{H}^2\left( \frac{\rho_\text{m}}{\rho_\text{tot}} \delta_\text{m} + \frac{\rho_\text{r}}{\rho_\text{tot}} \delta_\text{r} \right)}{3\mathcal{H}},
	\label{eq:phidot_eq}
\end{align}
where $\rho_\text{tot} = \rho_\text{m} + \rho_\text{r}$.

\begin{figure}[ht] 
        \centering \includegraphics[width=0.7 \textwidth]{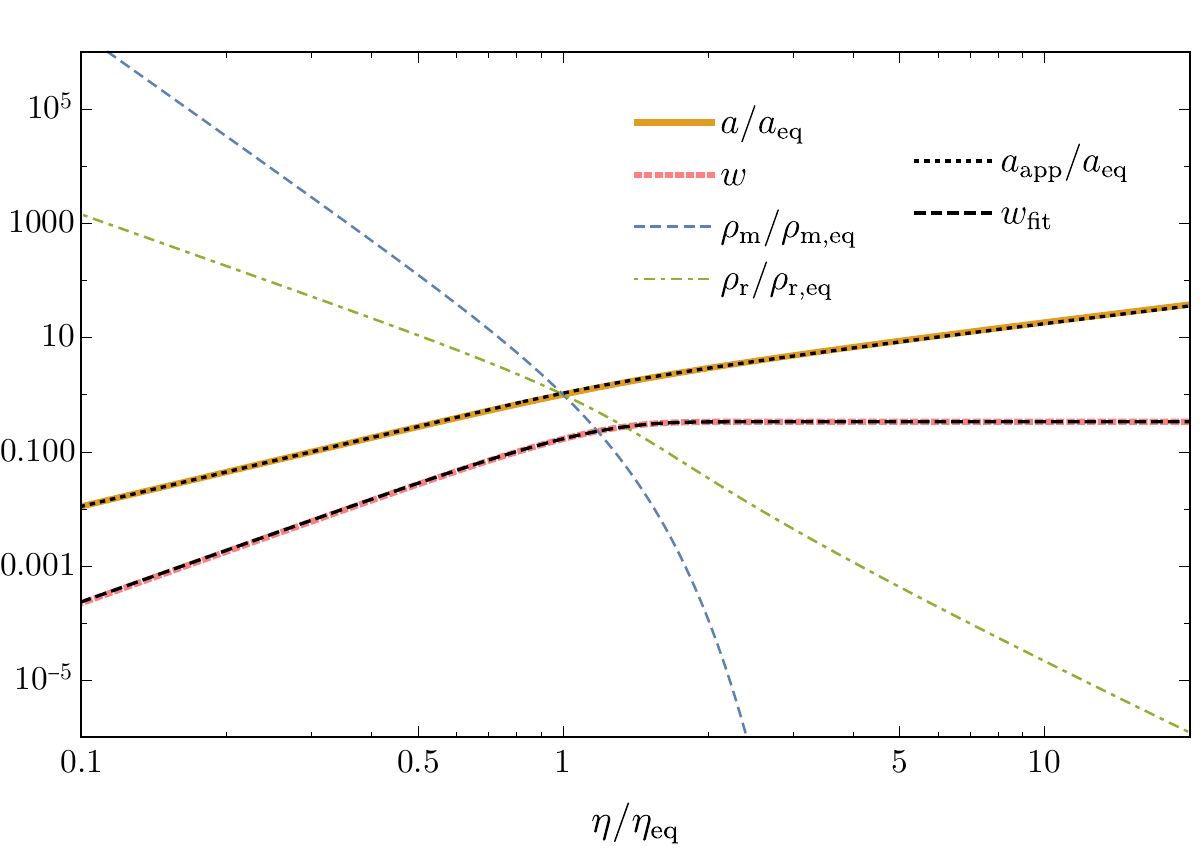}
        \caption{
  Time dependences of the scale factor, the energy densities, normalized by their values at $\eta = \eta_\text{eq}$, and the equation-of-state parameter.
  $a_\text{app}$ is given by Eq.~(\ref{eq:a_app}) with $\eta_\R = 0.83\eta_\text{eq}$ and $w_\text{fit}$ is given by Eq.~(\ref{eq:w_fit}).  Note that $\rho_{\text{m},\text{eq}}=\rho_{\text{r},\text{eq}}$ by definition.
        }
        \label{fig:back_summary}
\end{figure}

Figure~\ref{fig:back_summary} shows the numerical results for the evolutions of the background quantities, such as $a$, $\rho_\text{m}$, $\rho_\text{r}$, and $w$.
We define $\eta_\text{eq}$ as the conformal time when $\rho_\text{m} = \rho_\text{r}$. Note that there is one-to-one correspondence between $\Gamma$ and $\eta_{\mathrm{eq}}$ and hence this and the subsequent figures do not depend on the specific choice of $\Gamma$. 
We also plot the approximation formula  
Eq.~(\ref{eq:a_app}) with $\eta_\R=0.83 \eta_\text{eq}$, as well as the following fitting formula:
\begin{align}
w_\text{fit} = \frac{1}{3} \left( 1-\exp \left(-0.7 \left(\frac{\eta}{\eta_\text{eq}}\right)^3 \right) \right).
\label{eq:w_fit}
\end{align}
Both the formulas fit the numerical results very well.
We will explain the reason why we take $\eta_\R = 0.83 \eta_\text{eq}$ shortly. The fact that the approximation formula for the scale factor fits the numerical result well may indicate that using the exact solutions for the Green's functions during an eMD era and the RD era is a good approximation, noting the Green's functions are determined by the scale factor (see Eq.~\eqref{eq:g_formula}).

\begin{figure}[ht] 
        \centering \includegraphics[width=0.7 \textwidth]{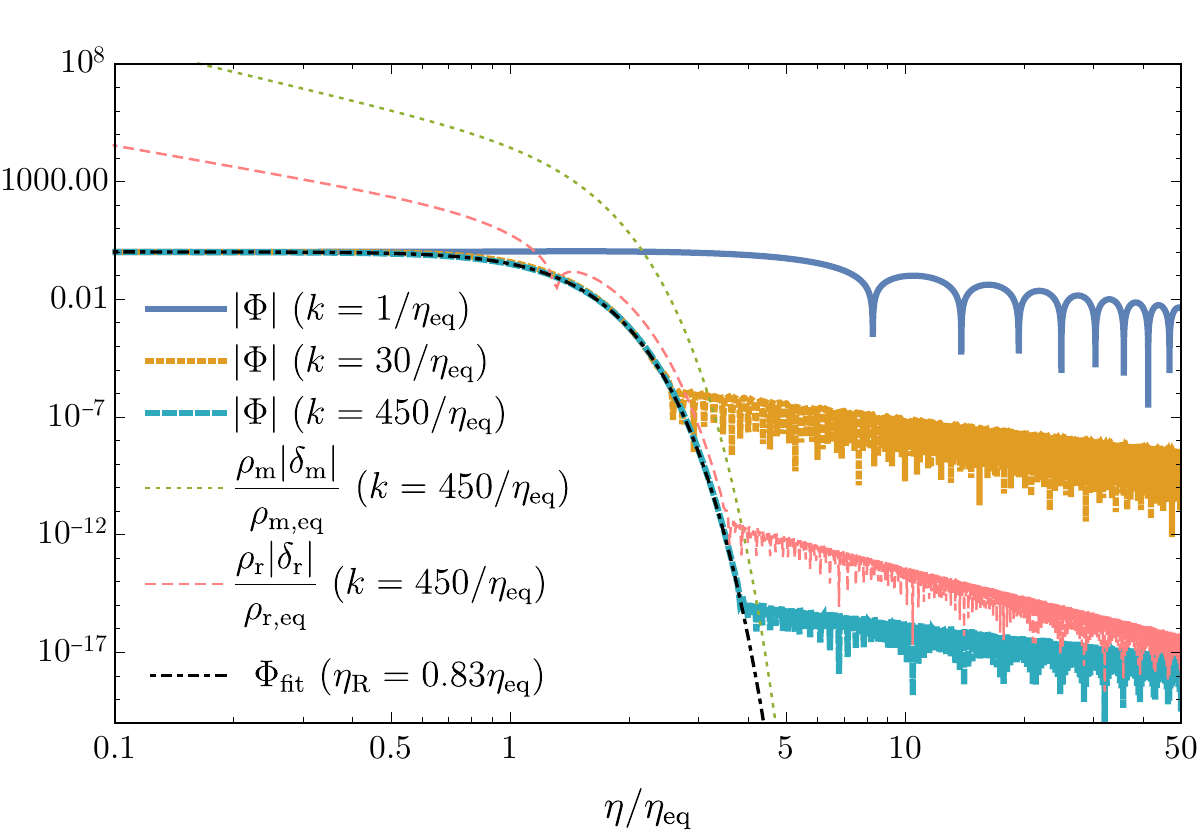}
        \caption{
  Time dependences of the gravitational potential and the energy density perturbations.  $\Phi_{\text{fit}}$ is given by Eq.~\eqref{eq:phi_app_f}. 
        }
        \label{fig:pertb_summary}
\end{figure}

Figure~\ref{fig:pertb_summary} shows the numerical results of the evolutions of the perturbations.
Here, we have assumed the following adiabatic initial conditions \cite{Mukhanov:991646}:\footnote{
	Strictly speaking, in Ref.~\cite{Mukhanov:991646}, the initial conditions for $\theta_{\text{m}/\text{r}}$ are not discussed.
	Howerver, we can easily derive the initial conditions for them substituting the initial condition for $\delta_{\text{m}}$ into Eq.~(\ref{eq:theta_m}).
}
\begin{align}
&\delta_{\text{m},\text{ini}} = -2\Phi_\text{ini}, \ \delta_{\text{r},\text{ini}} = \frac{4}{3} \delta_{\text{m},\text{ini}}, \ \theta_{\text{m},\text{ini}} = \theta_{\text{r},\text{ini}} = \frac{k^2 \eta}{3} \Phi_\text{ini}.
\label{eq:ini_condition}
\end{align}
Note that we study the linear regime, and thus the overall
normalization of perturbations does not matter in the figures, hence
we take $\Phi_{\text{ini}}=1$, or equivalently we plot the transfer
function.  In Fig.~\ref{fig:pertb_summary}, we can see that for
perturbation modes that entered the horizon well before the
transition ($k \gg 1/\eta_\text{eq}$), the gravitational potential
$\Phi$ exponentially decays soon after the equality time, and after a while,
$\Phi$ starts to oscillate due to radiation pressure, with the
amplitude decaying less rapidly ($\propto \eta^{-2}$).

\begin{figure}[ht] 
        \centering \includegraphics[width=0.7 \textwidth]{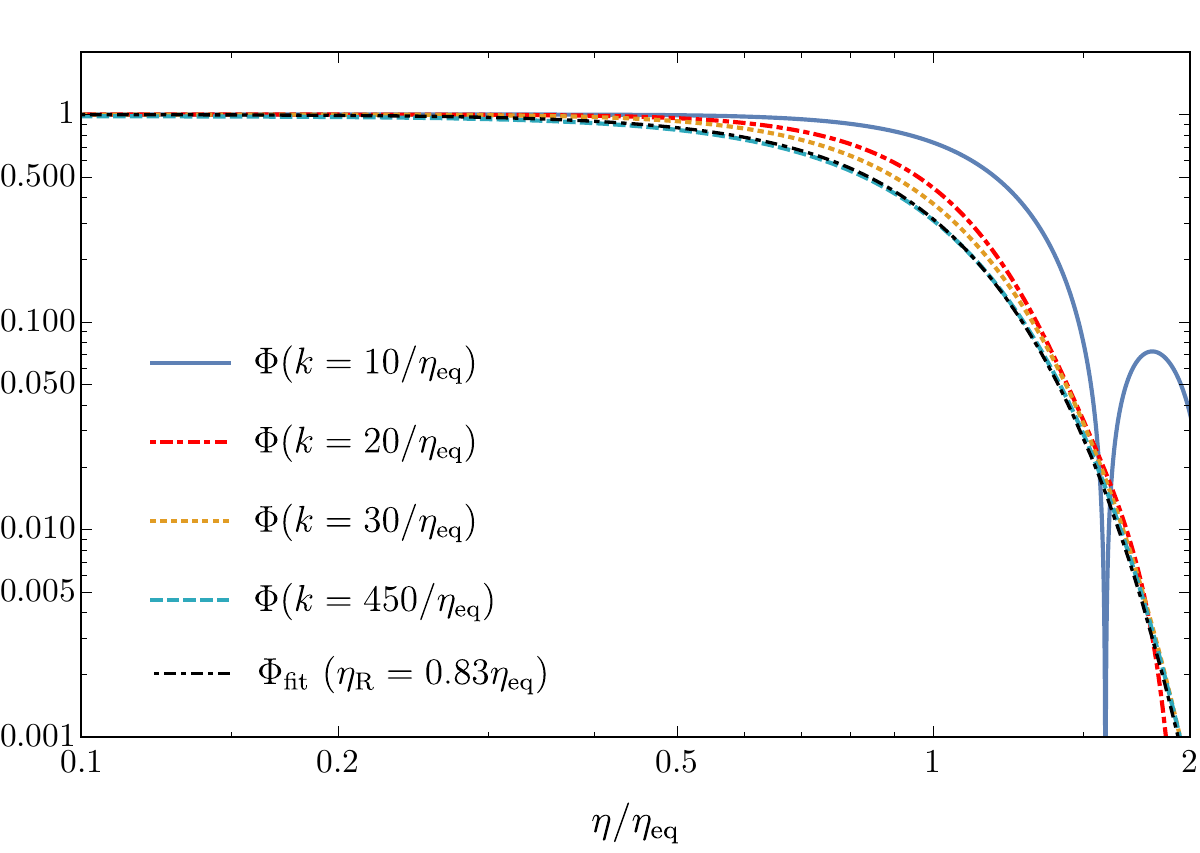}
        \caption{
        Evolutions of $\Phi$ with different wave numbers ($10/\eta_\text{eq} \leq k \leq 450/\eta_\text{eq}$) and the fitting formula [Eq.~\eqref{eq:phi_app_f}].
        }
        \label{fig:phi_evo}
\end{figure}

Here, we explain how to derive an approximation formula of $\Phi$ which
describes its exponential decay. The formula will be used to calculate the
induced GWs. First, from Eq.~(\ref{eq:phidot_eq}), $\Phi$ can be
approximated to be
$k^2\Phi \simeq \frac{3}{2} \mathcal H^2 \left(
  \frac{\rho_\text{m}}{\rho_\text{tot}} \delta_\text{m} +
  \frac{\rho_\text{r}}{\rho_\text{tot}} \delta_\text{r} \right)$
in the subhorizon limit.  For modes with $k\gg 1/\eta_\text{eq}$,
during $1/k \ll \eta \ll \eta_\text{eq}$, $\delta_\text{m}$ grows but
$\delta_\text{r}$ does not. Therefore, even after $\eta_\text{eq}$, the
evolution of $\Phi$ is dominated by $\rho_\text{m} \delta_\text{m}$ for a while.
During this phase, $\rho_\text{m}$ decays exponentially and then we can expect
$\Phi$ is proportional to $\rho_\text{m}$.  Radiation density perturbations
$\rho_\text{r} \delta_\text{r}$ also decay following the decay of $\Phi$ around this
phase.  After a while, the evolution of $\Phi$ is dominated by
$\rho_\text{r} \delta_\text{r}$ and then $\Phi$ starts to oscillate.  Neglecting
this radiation term and the expansion of the Universe during the
transition for simplicity, we can approximate $\Phi$ as
\begin{align}
\label{eq:phi_app_f}
\Phi  &\sim \exp \left(-\int^\eta \dd \bar \eta a(\bar \eta) \Gamma \right) \nonumber \\
& = \begin{cases} 
\exp \left( -\frac{2}{3} \left(\frac{\eta}{\eta_\R} \right)^3 \right) & (\eta < \eta_\R), \\
\exp\left( -2 \left( \left(\frac{\eta}{\eta_\R}\right)^2 - \frac{\eta}{\eta_\R} + \frac{1}{3} \right) \right) & (\eta \geq \eta_\R),
\end{cases}\\
&\equiv \Phi_\text{fit}, \nonumber
\end{align}
where we have used Eq.~(\ref{eq:a_app}) and assumed
$a(\eta_\R) \Gamma = \mathcal H(\eta_\R) (= 2/\eta_\R)$.  Note that
since we consider a gradual decay of the dominant field, there is some
ambiguity in the definition of $\eta_\R$, and hence we may choose
values of $\eta_\text{R}$ which are slightly different from
$\eta_\text{eq}$.  From Fig.~\ref{fig:pertb_summary}, we can see that
if we take $\eta_\R = 0.83 \eta_\text{eq}$, the approximation formula
fits the non-oscillating part of the numerical results very well.  As we have seen in
Fig.~\ref{fig:back_summary}, the approximation formula of the scale factor, given by
Eq.~(\ref{eq:a_app}) with $\eta_\R = 0.83 \eta_\text{eq}$, also fits
the numerical result very well.  Therefore we take
$\eta_\R = 0.83 \eta_\text{eq}$ when we use Eqs.~(\ref{eq:a_app}) and
(\ref{eq:phi_app_f}) in the following.  Figure~\ref{fig:phi_evo} is an
enlarged view of the evolution of $\Phi$ for several modes around the
transition.  From this figure, we can see that, for
$k \gtrsim 30/\eta_\text{eq}$, the exponential decay of $\Phi$ can be fitted
by Eq.~(\ref{eq:phi_app_f}).

\section{Calculations of induced gravitational waves}
\label{sec:gw_calculation}

In this section, we explain how we approximately calculate induced GWs in gradual-transition scenarios and present the results.
We consider the power spectrum given by 
\begin{align}
\mathcal P_\zeta = A\, \Theta(k-30/\eta_\text{eq}) \Theta(k_\text{max} -k),
\label{eq:pzeta}
\end{align} 
where $A$ represents the amplitude. We focus on modes with
$k>30/\eta_{\mathrm{eq}}$ so that we can use the fitting formula for
$\Phi$ obtained in the previous section. In addition, we have
introduced the cutoff scale $k_\text{max}$.  This cutoff scale
corresponds to the horizon scale at the start of an eMD era, or the
scale that is entering the non-linear regime at $\eta_{\mathrm{R}}$,
i.e., the amplitude of matter density perturbations on such a scale becomes unity at $\eta_\R$.
Since the formulas shown above are invalid in the non-linear regime,
we need to introduce this cutoff scale to limit our analysis to the
linear regime when there exist scales that enter the non-linear
regime before $\eta_{\mathrm{R}}$.  The scale corresponding to the
non-linear growth of the perturbations is roughly estimated
as~\cite{Assadullahi:2009nf, Kohri:2018awv}
\begin{align}
k_\text{NL} \sim \sqrt{\frac{5}{2}} \mathcal P_\zeta^{-1/4} \mathcal{H} (\eta_\R) \sim 470 /\eta_\R.
\label{eq:nl_scale}
\end{align}
Therefore our calculations are valid only for $k_\text{max} \lesssim 470/\eta_\R$. 
To make the differences from the previous works look clear, we take $k_\text{max} = 450/\eta_\R$ in the following.

With the above power spectrum, we can use the fitting formula of
$\Phi$, given in Eq.~(\ref{eq:phi_app_f}).  As we have seen in
Sec.~\ref{sec:phi_evo}, the expression of the scale factor for a
sudden transition, given in Eq.~(\ref{eq:a_app}), describes the
numerical results very well, and this justifies the use of the
formulas for the induced GWs, introduced in Sec.~\ref{sec:formalism} with the separation of the contributions at $\eta_{\mathrm{R}}$.
Then, to calculate induced GWs, we substitute the fitting formulas of
$w$ and $\Phi$ for a gradual transition, given in
Eqs.~(\ref{eq:w_fit}) and (\ref{eq:phi_app_f}), into
Eq.~(\ref{eq:f_def}), thereby taking into account the graduality of
the transition.  Note that we neglect contributions from the
oscillation phases of $\Phi$ because the oscillation amplitude is very
small for $k>30/\eta_\text{eq}$, noting that the power spectrum of
the induced GWs is basically proportional to the fourth power of
$\Phi$. For example, in Fig.~\ref{fig:pertb_summary}, we can see that
$\Phi$ with $k=30/\eta_\text{eq}$ starts to oscillate with the
amplitude $\Phi \sim \mathcal O(10^{-6})$.

\begin{figure}[ht] 
        \centering \includegraphics[width=0.7 \textwidth]{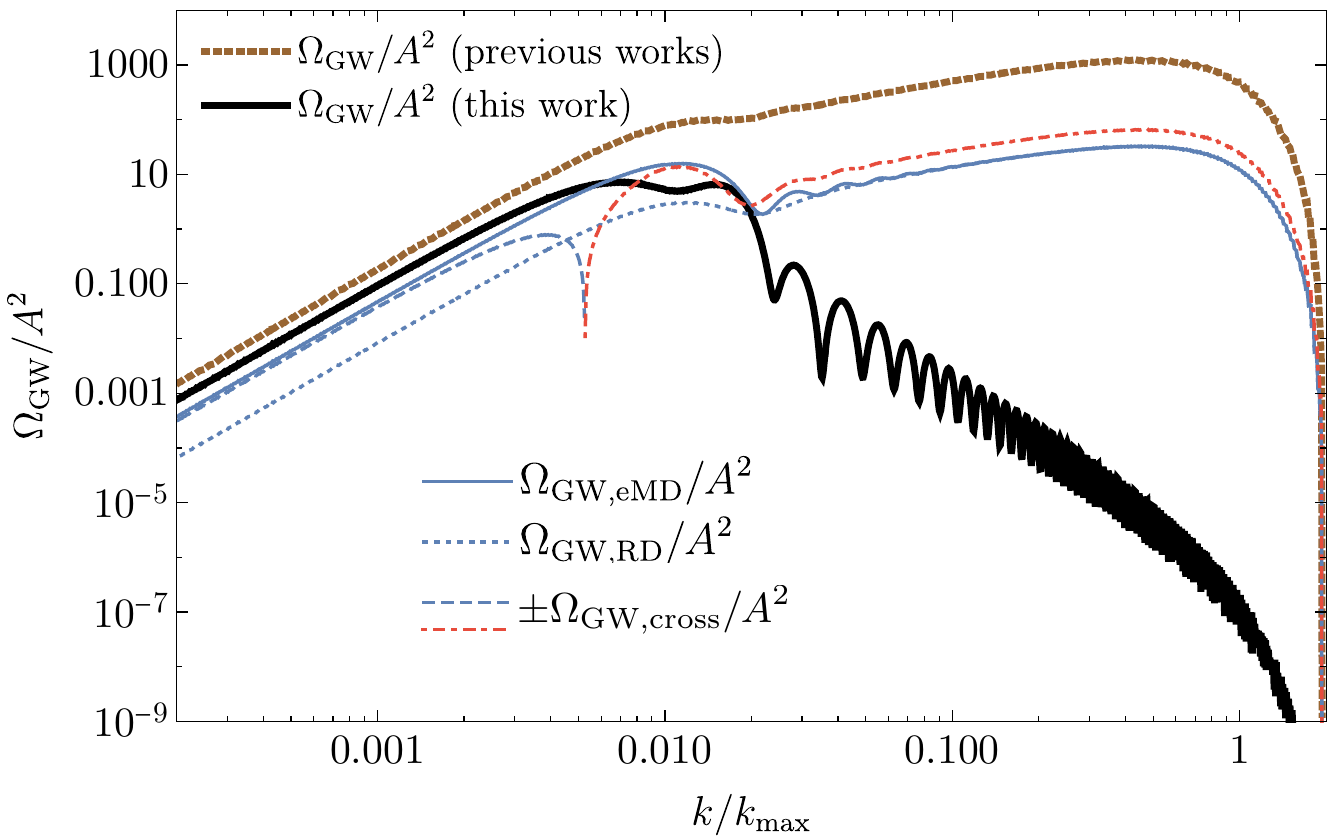}
        \caption{
         Spectrum of induced GWs at $\eta = \eta_c$ associated with a gradual transition from an eMD era to the RD era.
        For all the plots, $k_\text{max}= 450/\eta_\R$ is assumed and the primordial power spectrum used is given in Eq.~(\ref{eq:pzeta}).
        Our results for $\Omega_\text{GW}$ are shown by the thick black solid line.  It is the sum of each component: $\Omega_{\text{GW,eMD}}$ (blue solid), $\Omega_{\text{GW,RD}}$ (blue dotted), and $\pm\Omega_{\text{GW,cross}}$ (blue dashed/red dot-dashed).
        The thick brown dotted line shows the result using Ref.~\cite{Kohri:2018awv}, additionally taking into account the factor $1/4$ mentioned in footnote~\ref{fn:footnotename}, with the same assumptions as in Refs.~\cite{Assadullahi:2009nf}, under which $\Phi=1$ until $\eta = \eta_\R$ and there is no GW production for $\eta > \eta_\R$ ($I_\text{RD} =0$).
        Reduction from the brown dotted line to the black solid line shows the effects of the gradual transition.
        }
        \label{fig:gw_sup_result}
\end{figure}

Figure~\ref{fig:gw_sup_result} shows the numerical results for the
induced GWs.  From this figure, we can see that the induced GWs (thick black solid line) are
suppressed compared to those derived with the setups in the previous
works (brown dotted line).  In this figure, each component of the induced GWs is also
shown (blue solid, blue dotted, and blue dashed/red dot-dashed lines for $\Omega_\text{GW,eMD}, \Omega_\text{GW,RD}$, and $\pm \Omega_\text{GW,cross}$, respectively).  From these plots, we  understand two reasons for the
suppression.  First, each component of the induced GWs is suppressed
because $\Phi$ is smaller than unity around the transition, as we have
seen in the previous section.  The second reason is that a
cancellation occurs between those components.  (Note again that
$\Omega_\text{GW,cross}$ can be negative by definition.)

\begin{figure}[ht] 
        \centering \includegraphics[width=0.7 \textwidth]{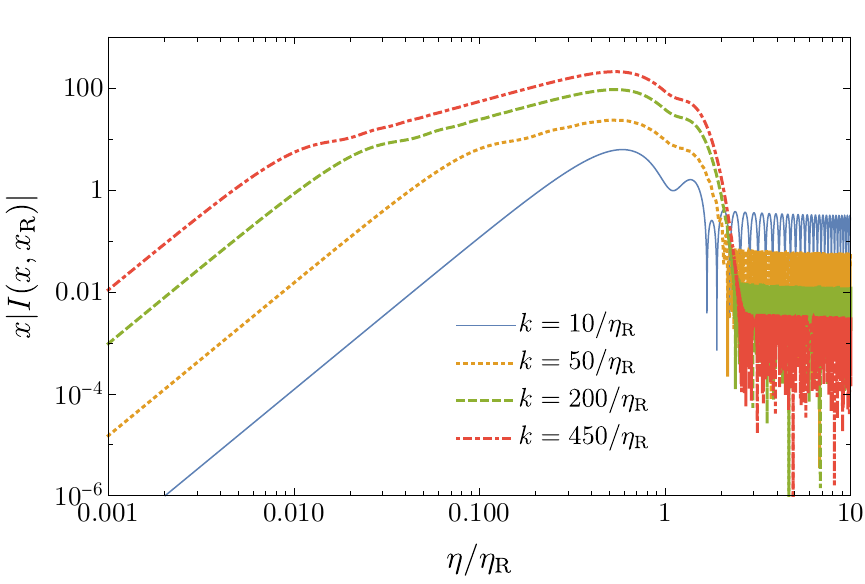}
        \caption{
        Time dependences of $x |I(x,x_\R)|$, defined in Eq.~(\ref{eq:i_formula}).
        We omit the arguments $u$ and $v$ because we approximate $\Phi$ as $\Phi_\text{fit}$, defined in Eq.~(\ref{eq:phi_app_f}), and hence $I$ does not depend on $u$ or $v$.
        }
        \label{fig:i_compared}
\end{figure}

To see how the cancellation arises, we show the time dependence of $x |I|$ in Fig.~\ref{fig:i_compared}.
For $\eta > \eta_\R$, we use Eq.~(\ref{eq:i_formula_emd1}) and, for $\eta< \eta_\R$, we use 
\begin{align}
I(u,v,x,x_\R) = \int^{x}_0 \dd \bar{x}  \left( \frac{\bar x}{x} \right)^2 k G_k^{\text{eMD}}(\eta, \bar \eta) f(v,u,\bar x,x_\R),
\label{eq:i_formula_emdonly}
\end{align}
which can be derived from Eq.~(\ref{eq:i_formula}), with $k G_k^{\text{eMD}}$ given by~\cite{Baumann:2007zm, Assadullahi:2009nf}
\begin{align}
	kG_k^\text{eMD} (\eta, \bar \eta) = - x \bar x (j_1(x) y_1(\bar x) - y_1(x) j_1(\bar x)).
\label{eq:gk_emd}
\end{align}
From Fig.~\ref{fig:i_compared}, we can see that $x I$ grows when
$\eta \ll \eta_\R$, but around the transition ($\eta \sim \eta_\R$),
$x I$ stops growing and starts to decrease due to the decay of $\Phi$.
This again shows the cancellation between $I_\text{eMD}$ and
$I_\text{RD}$.  After the transition ($\eta > \mathcal O(1) \eta_\R$),
$x I$ oscillates with its amplitude being almost constant. 
Since the evolutions of $I$ correspond to those of the tensor perturbations, the behavior of $x I$ can be interpreted as follows. 
During an eMD era, since the source term in Eq.~(\ref{eq:h_eom}) is almost constant, the amplitude of the tensor perturbation is given as $h^\lambda_{\bm{k}} \simeq 4 S^\lambda_{\bm{k}}/k^2$ in the subhorizon limit~\cite{Baumann:2007zm,Assadullahi:2009nf}. Here, we consider a gradual transition from an eMD era to the RD era and therefore the amplitude of the tensor perturbations decays on subhorizon scales, following the gradual decay of the source during the transition, which corresponds to the decay of $x I$ around $\eta \sim \eta_\R$. After a while, the tensor perturbations decouple from the source and behave as freely propagating GWs, which corresponds to the oscillation of $x I$ for $\eta > \mathcal O(1) \eta_\R$.

Note that, also to obtain
Fig.~\ref{fig:i_compared}, we use the approximation formula $\Phi_\text{fit}$
for $\Phi$ and neglect the oscillation behavior of $\Phi$ after its
exponential decay.  This also implies that the oscillation behavior of $x I$
in Fig.~\ref{fig:i_compared}  (already
present before $\eta = 2.5\eta_\text{eq} \simeq 3.0 \eta_\text{R}$) has nothing to do with the oscillation
behavior of $\Phi$ in Fig.~\ref{fig:pertb_summary} (appearing
after $\eta = 2.5 \eta_\text{eq} $ in $k > 30/\eta_\text{eq}$), and the
contributions from the oscillations of $\Phi$ can be neglected because
of its small amplitude if we consider $k> 30/\eta_\text{eq}$, as we
have already mentioned.

\section{Discussion}
\label{sec:discussion}

In this paper, we have revisited the effects of an eMD era on the GWs
induced by the scalar perturbations.  We have considered a
case where the energy density in the eMD era is dominated by the field
decaying to radiation with a constant decay rate, which leads to a
gradual transition to the RD era.  We have taken into account the
evolution of the gravitational potential $\Phi$ during the transition
and found that the existence of an eMD era does not necessarily
increase the induced GWs as were reported in previous works.  This is due
to the exponential decay of $\Phi$ during the transition.

Throughout this work, to spotlight the effects of an eMD era on induced GWs, we have focused on the GWs induced by the perturbations entering the horizon during an eMD era.
However, we have neglected the contributions from the perturbations entering the horizon during an eMD era but relatively near the transition ($1/\eta_\R < k<30/\eta_\text{eq}$) to avoid high computational costs.
If we consider the scale-invariant spectrum of curvature perturbations, these perturbations could produce GWs in addition to the results in Fig.~\ref{fig:gw_sup_result}. 
We leave the analysis of these additional effects of an eMD era for our future work in a separate paper.
In addition, we have not considered the GWs induced by the perturbations entering the horizon much before the transition ($k>470/\eta_\R$).
If we consider the scale-invariant spectrum and a long-lasting eMD era, the perturbations on $k>470/\eta_\R$ could also induce additional GWs.
Such perturbations may have become non-linear during the eMD
era, and our formula based on the linear perturbation theory cannot be
applied.  Although there are works discussing GWs induced by
non-linear scalar perturbations in this
context~\cite{Jedamzik:2010dq,Jedamzik:2010hq}, there are still some
uncertainties about the predictions~\cite{Jedamzik:2010dq, Jedamzik:2010hq}. Therefore,
the amount of the induced GWs predicted in the current work can be
regarded as lower bounds on the total induced GWs.

Although we assume that the decay rate is constant throughout this
paper, if we consider a time-dependent decay rate causing the field to
decay much faster than the Hubble time at that time, the induced GWs
can be significantly enhanced.  We discuss this issue in
Ref.~\cite{Inomata:2019ivs}.


\acknowledgments 

The authors thank Matthew Pearce, Csaba Balazs, Lauren Pearce, and Graham White for pointing out the error in Figs.~\ref{fig:gw_sup_result} and \ref{fig:i_compared} in the previous version of this paper.
KK and TT thank Sachiko Kuroyanagi for useful discussions. 
KI acknowledges Misao Sasaki, Tomohiro Fujita, Teruaki Suyama, and Masahide Yamaguchi for useful comments.
KI and TN are grateful to KEK or hospitality received during this work. 
This work was supported in part by World Premier International Research
Center Initiative (WPI Initiative), MEXT, Japan, the JSPS Research
Fellowship for Young Scientists (KI and TT), JSPS KAKENHI Grants No.~JP18J12728 (KI), 
No.~JP17H01131 (KK), and No.~JP17J00731 (TT), MEXT KAKENHI Grants
No.~JP15H05889 (KK), No.~JP18H04594 (KK) and No.~JP19H05114 (KK), and
Advanced Leading Graduate Course for Photon Science (KI).





\end{document}